\documentclass[journal=langd5,manuscript=article]{achemso}
\usepackage{graphicx} 
\usepackage{amsmath}
\usepackage{xcolor}
\usepackage{soul,ulem}
\usepackage{physics}
\usepackage[version=3]{mhchem} 

\newcommand{\msrev}[1]{\textcolor{black}{#1}}

\title{
\textcolor{black}{Effect of simple shear on knotted polymer coils and globules}
}
\author{Andrey Milchev}
\email{milchev@ipc.bas.bg}
\affiliation{Bulgarian Academy of Sciences, Institute of Physical Chemistry, 1113 Sofia, Bulgaria}
\author{Maurice P. Schmitt}
 \affiliation{Johannes Gutenberg-Universit\"{a}t, Institut f\"{u}r Physik, Mainz, Germany}
\author{Peter Virnau}
 \affiliation{Johannes Gutenberg-Universit\"{a}t, Institut f\"{u}r Physik, Mainz, Germany}
\date{\today}

\begin{document}
\maketitle
 
\begin{abstract}

We explore the effect of \textcolor{black}{Couette} flow on 
knotted linear polymer chains with extensive Molecular Dynamics (MD) simulations. Hydrodynamic interactions are accounted for by means of Multi-Particle Collision Dynamics (MPCD). The polymer chain, containing originally a simple trefoil knot at rest, is described by a coarse-grained bead-spring model in a coil or globular state. 
We demonstrate that under shear existing loosely localized knots in polymer  coils typically {\it tighten} to several segments beyond a certain shear rate threshold. \textcolor{black}{At large shear rates the polymer undergoes a tumbling-like motion during which knot sizes can fluctuate.}
In contrast, sheared knotted globules unwind into a \textcolor{black}{convoluted} pearl-necklace structure of sub-globules \textcolor{black}{that folds back onto itself and in which knot types change over time.}
\end{abstract}

\section{Introduction}
\label{sect:intro}

Knots play an essential role in polymers \cite{Micheletti_PhRep2011} and biological molecules \cite{Knots_review}, most notably in DNA \cite{ArsuagaPNAS2002, ArsuagaPNAS2005, MarenduzzoPNAS2009, Plesa2016, KumarSharma2019} and to some degree also in proteins \cite{Taylor:Nature:2000, Virnau:PLoScb:2006, Potestio, Mallam:PNAS:2010, Boelinger_2010, Virnau_2011, Jarmolinska_2019, Brems_2022}. Just like in macroscopic strings or ropes, they occur spontaneously with increasing probability as the length of the polymer \cite{Frisch:JACS:1961, Delbruck_knot_62, Vologodskii:1974, Sumners:Whittington:JPA:1988} or its compactness increases \cite{Mansfield:Macro:1994, Kantor}. Even though knots are formally defined only in closed rings \cite{Adams:1994}, "open knots" referred to hereafter simply as knots, can unambiguously be described and detected by applying an appropriate closure \cite{Virnau, Tubiana:PTPS:2011}. Formation and removal of knots typically takes place by slithering motion of the polymer chains via the ends, e.g., in a polymer melt \cite{Foteinopoulou:PRL:2008, Meyer:ACSMacro:2018:knots_melt, Jianrui2020} and in confinement, and is enhanced (up to a certain point) by chain stiffness \cite{Coronel:SoftMatt:2017:bending}. 
But how do external forces acting on the polymer chain affect existing knots in it?
 
In cases when a protein has been pulled by forces applied to its ends, it has been found earlier that a knot may either get tightened \cite{Dzubiella} or untied \cite{Ohta}, depending on the way the pulling force (e.g., single-molecule Atomic Force Microscope) is applied. It has been shown \cite{Cieplak} that by pulling specific amino acids, one may retract a terminal segment of the backbone from the knotting loop and untangle the knot. While these studies have been based on applying a precisely chosen tensile force on selected segments along the chain backbone, the impact of common forces, ubiquitous in nature, as simple shear on knots remains insufficiently understood. 
\textcolor{black}{Knots have been studied in single ring polymers under shear with and without hydrodynamic interactions \cite{Liebetreu2018}. In these simulations, a tumbling motion was observed at higher shear-rates which resulted in a rather broad distribution of knot sizes. For linear chains} an investigation on dynamics and topology of knots in steady shear flow\cite{Kuei} has reported repeated unknotting-knotting transitions in flexible chains while Brownian dynamics simulations on the behavior of knotted chains in 
elongational flow \cite{Doyle} have concluded earlier that knots untie. \textcolor{black}{Other studies have exposed knotted polymers to AC/DC fields thus driving the knots \cite{diStefano:Softmatter:2014}.} 
In recent years, relaxation \cite{Doyle2}, dynamics \cite{Klotz:PRL:2018motion}, attraction \cite{Klotz_2020} and untying \cite{Soh_2018} of DNA knots stretched by elongational fields have also been studied experimentally.  
In contrast to knots, the impact of shear on the conformational and dynamic evolution of polymer chains themselves has been subject of intensive investigations for several decades now \cite{Tan,Smith,Lyulin,Kong,Baig,Nikoubashman} and is currently much better understood.

\section{Model}
\label{sect:model}

We use a coarse-grained description for the segments, representing a polymer chain molecule by a bead-spring model, i.e., a sequence of $N$ effective spherical monomeric units held together by suitable effective potentials. Any pair of beads $(i,j)$ in the system at distance $r =|\vec{r}_i - \vec{r}_j|$ interacts with the purely repulsive Weeks - Chandler - Andersen (WCA) potential \cite{WCA},
\begin{eqnarray}
U^{WCA}(r) &=& \begin{cases} 4 \epsilon [(\sigma /r)^{12}-(\sigma/r)^6+1/4], \; r \leq r_c =  2^{1/6} \sigma,\\
 0, \; r > 2^{1/6} \sigma   .
\end{cases} \label{eq1}
\end{eqnarray}
Here, $\epsilon$ sets the scale for the strength of this potential (in units of the thermal energy $k_BT$), $\sigma$ is the range, and we choose units of energy and length such that $\epsilon = 1$ and $\sigma = 1$. Neighboring monomer units along the chain experience the finitely extensible nonlinear elastic (FENE) potential \cite{KG},
\begin{equation} \label{eq2}
U^{\rm FENE} (r) = \left \{
\begin{aligned} &-0.5 k r^2_0 \ln \left [ 1-\left (\frac{r}{r_0} \right)^2 \right ], && r<r_0,\\
&\; \infty, &&r \ge r_0, \end{aligned} \right .
\end{equation}
with the standard choice of parameters $r_0 = 1.5\sigma$ and $k = 30\epsilon / \sigma^2$. Eqs. (\ref{eq1},\;\ref{eq2}) cause a rather sharp minimum for the bond length $\ell_b \approx 0.970~\sigma$, so that the root mean square fluctuation of the bond length is only of order $0.03$, irrespective of chain stiffness. In our investigation we study fully flexible polymer chains comprised of $N=512$ or $N=1024$ segments. While Eq.~\ref{eq1} describes the case of a swollen coil, for the globular state the cutoff was adjusted to $r_c=2.5\sigma$ \textcolor{black}{between all monomers} to allow for attractive interactions. Temperature $T=1~\epsilon/k_B$ was used for all simulations.
Chains are confined in a slit with smooth repulsive walls (using Eq.(\ref{eq1})) parallel to the $X$-direction and perpendicular to the $Z$-axis. The size of the simulation box is $L_x \times L_y \times L_z = 200 \times 80 \times 80$, or $200 \times 160 \times 160$, respectively, for the chains with $N=512$ and $N=1024$. For the integration of the equations of motion of the chain monomers we use the NVE-ensemble with a time step $dt_{MD} = 0.00025\tau_{MD}$ whereby the MD time unit is $\tau_{MD} = \sqrt{m\sigma^2/\epsilon}$. 

In MPCD, solvent particles are modeled as ideal point particles with unit mass $m=1$, and their motion is governed by alternating collision and streaming steps \cite{Malevanets,Gompper}. During the streaming step the solvent particles move ballistically within a time interval  $dt_\mathrm{MPCD} = 0.0025\tau_{MD}$. In the collision step all solvent particles are first sorted into cubic cells of volume $\sigma^3$ (we use density $\rho=5\sigma^{-3}$ of particles per cell), whereby particles within the same cell exchange momentum through stochastic collisions while conserving momentum on both the cellular and global level. Collisions are performed with a period of $10\times dt_{MPCD}$. In the present study we used the HOOMD-Blue $v2.9.4$ software package\cite{Travesset} on graphics processing units (GPUs) with an Andersen thermostat (AT) collision scheme \cite{Padding} which acts as the thermostat in our simulations. Before each collision phase, cells are shifted by a random three-dimensional vector with components drawn uniformly from $[-\sigma/2, \sigma/2]$ and MPCD-particle velocities are rotated by an angle of $\alpha = 130^\circ$ with respect to the center-of-mass velocity of the cell to ensure Galilean invariance\cite{Ihle}. To emulate Couette flow we create a constant velocity gradient along the $Z$-direction perpendicular to the direction of motion along the $X$-axis by fixing the velocity $\pm v$ of the walls at $\pm L_z/2$. At the channel (slit) walls we use {\it no-slip} boundary conditions imposed by a bounce-back rule at the walls. 
%


In order to present our data in terms of hydrodyamic numbers, such as the Weissenberg number, $\mathrm{Wi} = \dot{\gamma} \tau_0$, where $\tau_0$ is the longest relaxation time of the polymer chain, $\tau_0$ is determined from the exponential decay of the auto-correlation function, 
\begin{equation}\label{eq:corr}
  C(t) = \frac{\langle \vec{R}(t_0+t)\vec{R}(t)\rangle}{\langle \vec{R}(t_0)\vec{R}(t_0)\rangle} =\exp(-t/\tau_0) ,
\end{equation}  
 where $\vec{R}(t)$ is the end-to-end vector of the chain and $t_0$ is the reference time. For a \msrev{coil} with length $N=512$ we thus get $\tau_0 = 31892\, \tau_{MD}$, \msrev{and for $N=1024$ we get $\tau_0 = 214285.5\, \tau_{MD}$} cf. Eq.(\ref{eq:corr}).  
 
Note that in the swollen coil state, chances of observing a knot for chains of these sizes are typically in the order of 1$\%$ or less, so they are in a sense rare events \cite{Kantor}. At the same time they are weakly localized, i.e., they only occupy a fraction of the chain as indicated in Fig.~\ref{fig:snapshot}(a). The globular state on the other hand is frequently knotted at these lengths (approximately $40\%$ for $N=512$ and $80\%$ for $N=1024$) and knots are typically spread out all over the globule \cite{Kantor}.

During the simulation we monitor position and size of the existing knot (a simple trefoil knot \textcolor{black}{initially chosen from an equilibrium conformation} \msrev{with a size close to the most common equilibrium size and neither start nor end close to the chain ends} as depicted in Fig.~\ref{fig:snapshot}(a) for example). Knot detection is performed by computing the Alexander polynomial \cite{Adams:1994} after closure. Here, we apply a custom code\cite{Virnau} implementing the center of mass closure:  
Two (typically outward-pointing) half lines emerging from the termini along the direction from the center of mass through the respective ends are connected far away from the polymer.
Knot sizes are determined by successively removing beads from each end and repeating the respective closure procedure until the knot disappear. \textcolor{black}{For the analysis of globules in Couette flow, we have used the KymoKnot package \cite{Tubiana:EPJE:2018:kymoknot} which applies the minimally-interfering closure \cite{tubiana2012equilibrium}, as it allows for reliable detection of knot sizes even for complex knots. An interesting alternative in this context might be to rely on knot detection schemes for open chains without prior closure, which have been discussed recently e.g. in \cite{klotz2024revisiting,Herschberg_2023}.}
The conformational properties of the polymer chain, among others its mean squared end-to-end distance $\langle R^2 \rangle$, and mean squared radius of gyration $\langle R_g^2 \rangle$, describe the coil transformation into a stretched string of segments subject to applied shear, as well as the globule transformation into pearl-necklace conformations.

\section{Results}
\label{sect_res}
\subsection{Coil in Couette flow}

\begin{figure}[ht]
\begin{minipage}{1.0\linewidth}
\hspace{1.5cm}
 \includegraphics[width=0.33\textwidth]{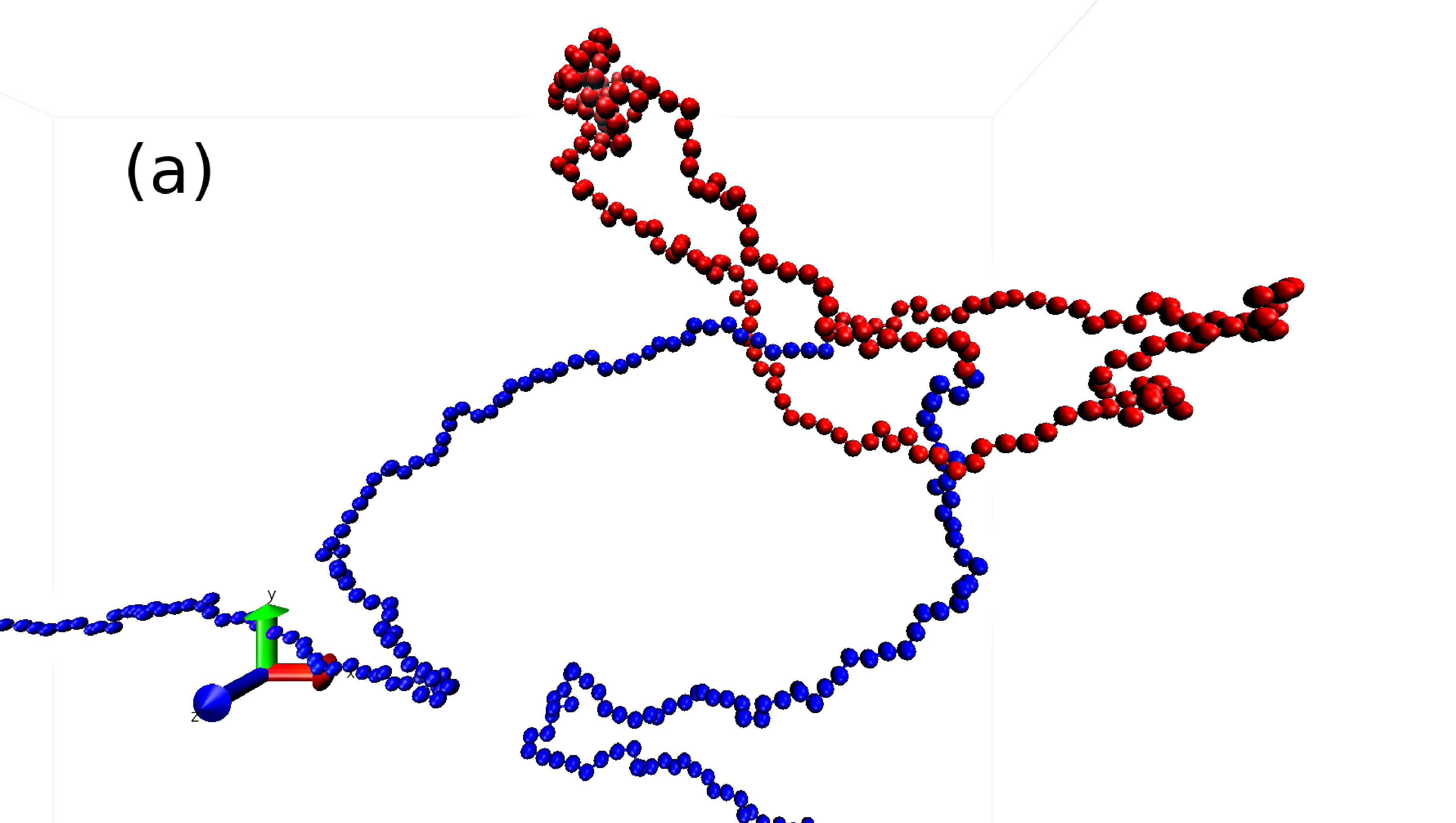}
\hspace{1.0cm}
\includegraphics[width=0.33\textwidth]{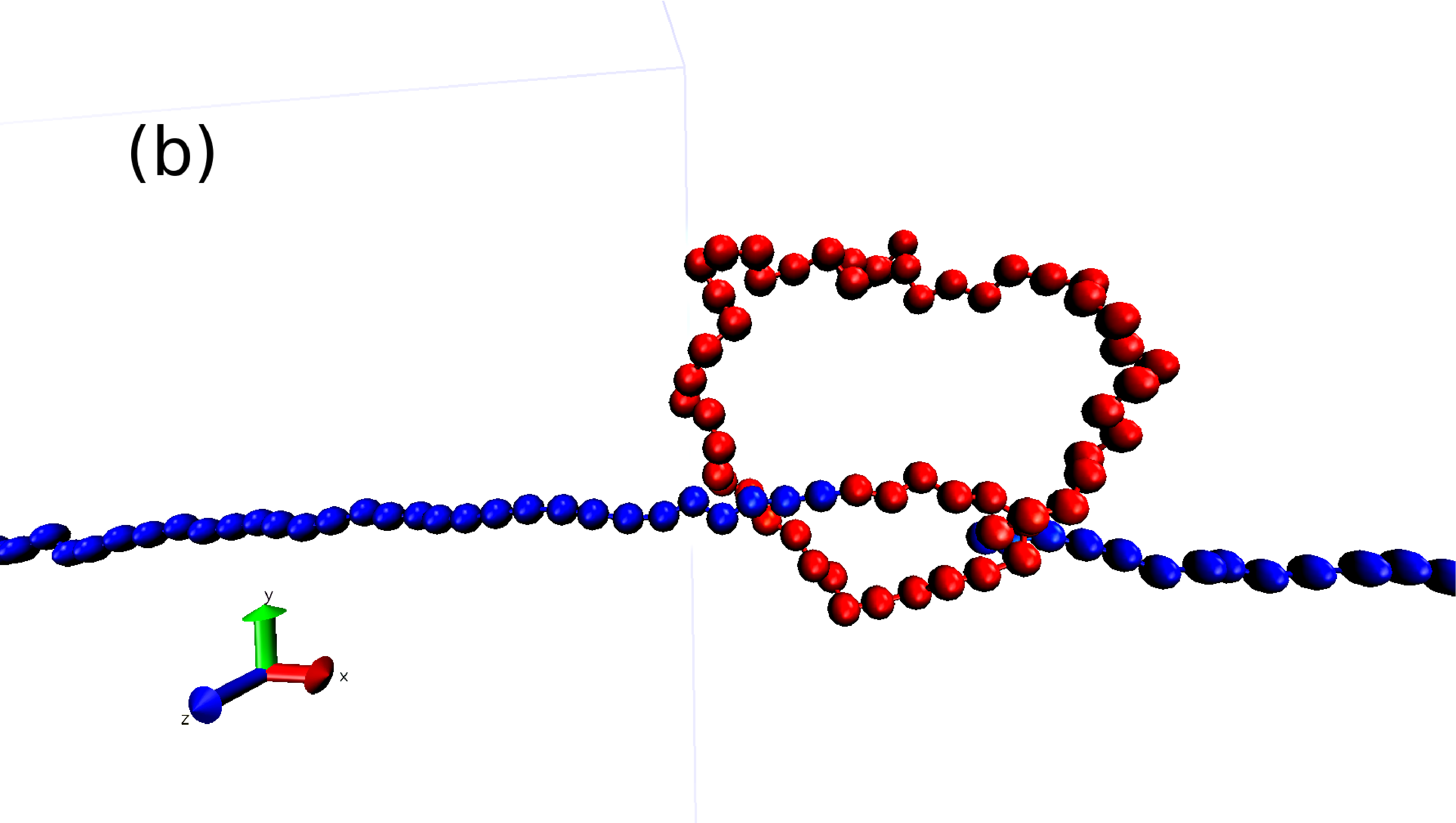}

\hspace{1.5cm}
 \includegraphics[width=0.33\textwidth]{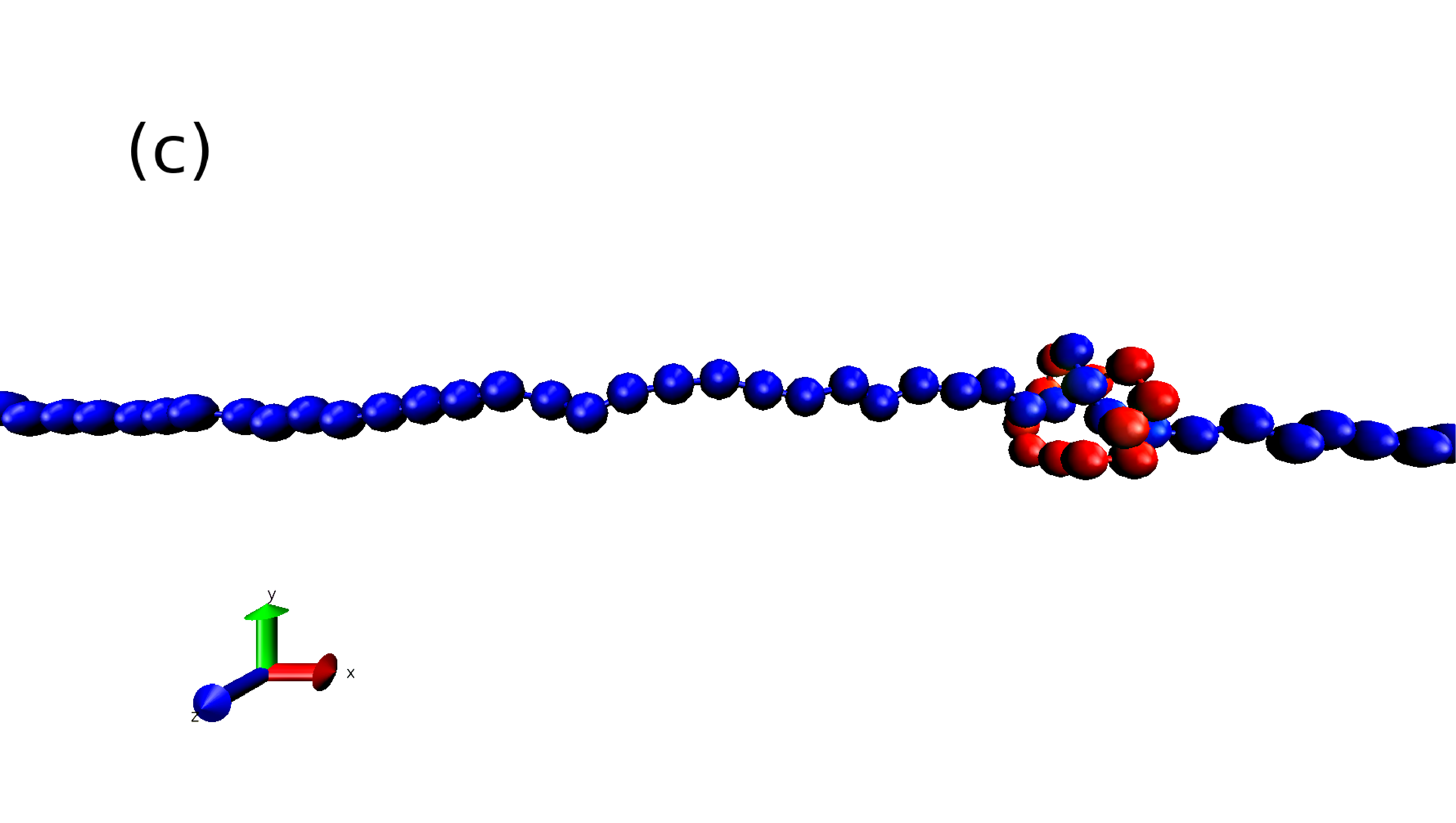}
\hspace{1.0cm}
 \includegraphics[width=0.33\textwidth]{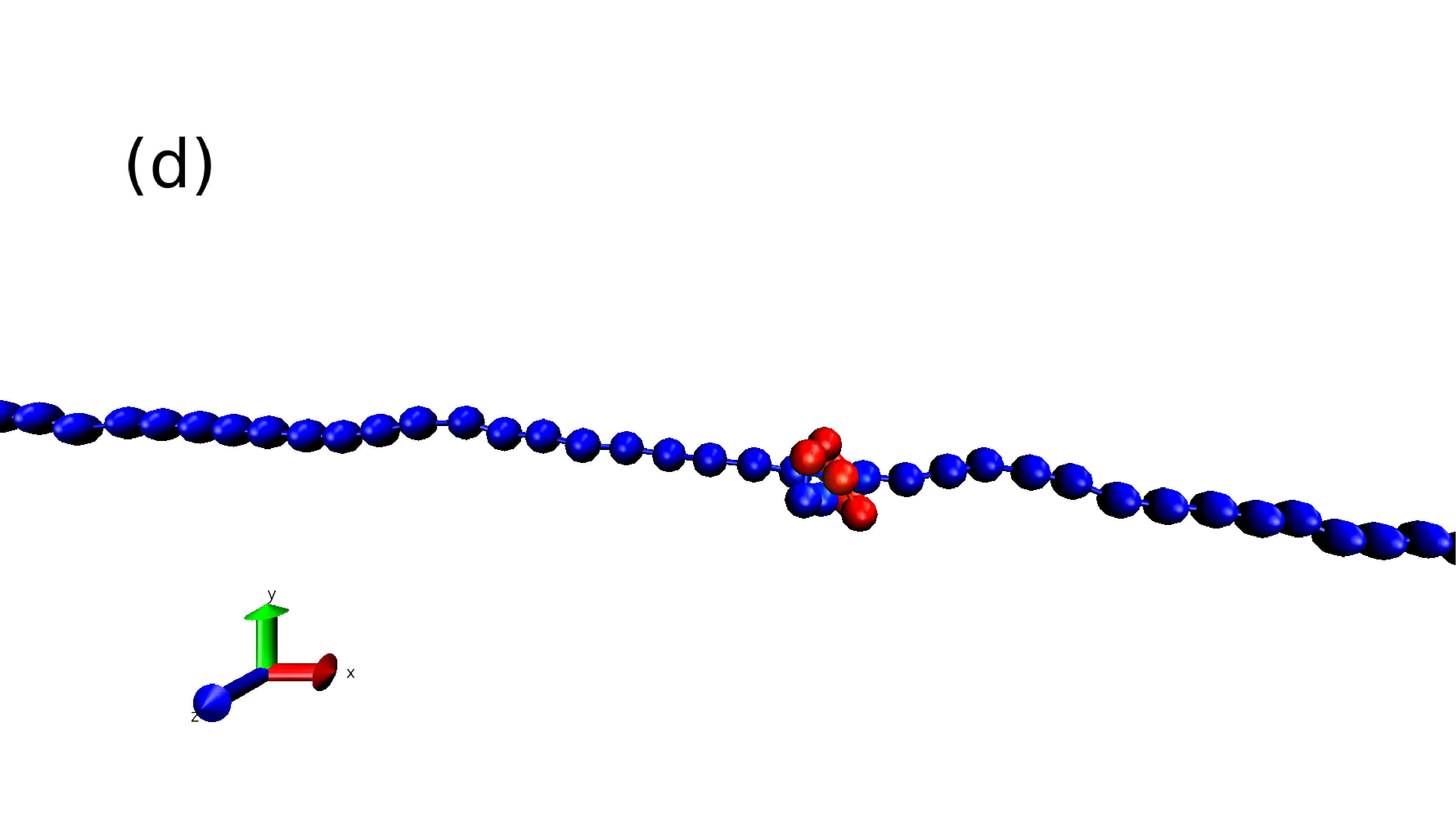}
\end{minipage}
 \caption{Time evolution of the trefoil knot of size $K_n(t)$ (in red) in a chain with $N=1024$ segments at shear rate $\dot{\gamma}=0.005$ \msrev{(Wi $\approx 1071$)} in Couette flow. Thus,  $K_n(55\tau_{MD})=186$ (a), $K_n(176\tau_{MD})=58$ (b), $K_n(200\tau_{MD})=12$ (c) and $K_n(775\tau_{MD})=7$ (d). The knotted section (as determined by our automated procedure) is depicted in red. All plots are prepared with VMD \cite{VMD}.
 }   
\label{fig:snapshot}
\end{figure}
Fig.\ref{fig:snapshot} shows a typical decrease of the knot size, 
$K_n(t)$, and its tightening with time $t$ after the onset of shear in the channel. Concomitantly, the coil itself turns progressively into a stretched string for the case of simple shear (Fig.\ref{fig:snapshot}b). Depending on the shear rate $\dot{\gamma}$, the knot size $K_n$ tends eventually to a minimal value of $5 - 15$ segments \textcolor{black}{at high shear} whereby the rate of tightening accelerates with growing shear rate \msrev{or Weissenberg number Wi} (Fig.\ref{fig:snapshot}c,d). 

\begin{figure}[ht!]
\hspace{0.0cm}

\begin{minipage}{1.\linewidth}


\includegraphics[width=.9\linewidth]{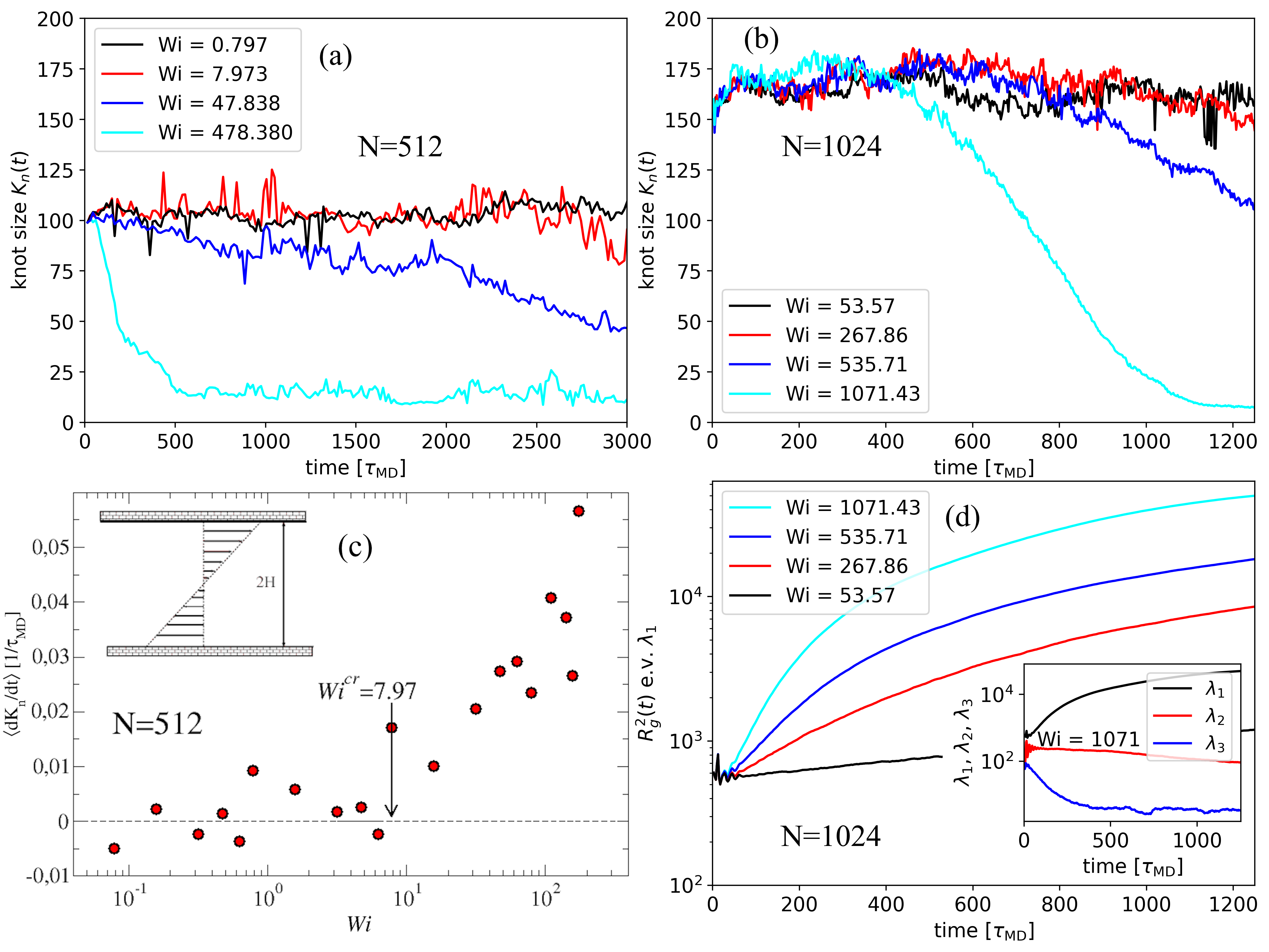}

 \caption{(a) \msrev{Average knot} size $K_n$ decreasing with time $t$ in a chain with $N=512$ segments for several shear rates\msrev{, or Weissenberg numbers $\mathrm{Wi}$}. (b) The same as in (a) for $N=1024$\msrev{.} 
 (c) \msrev{Onset of knot tightening at critical shear rate $\mathrm{Wi}^\mathrm{cr} \approx 7.97$ for chain length $N=512$.}  
 (d) Largest eigen-value of the mean-squared gyration radius $\langle R_g^2(t)\rangle$ for a chain with $N=1024$ and several \msrev{Weissenberg numbers.} 
 The inset shows the variation of the eigenvalues of the tensor $\langle R_g^2\rangle$: $\lambda_1(t)  > \lambda_2(t) > \lambda_3(t)$, indicating the transformation of the coil into a stretched string with time for \msrev{$\mathrm{Wi} \approx 1071$}. }  \label{fig:Rg2}
 \end{minipage}
\end{figure}

\textcolor{black}{The trefoil knot length evolution over time goes from approximately random fluctuations at very low shear rates, to tightening at larger shear (Figs.~\ref{fig:Rg2}(a,b)), which can also be observed for coils in Poiseuille flow (see SI). For the longer chain, $N=1024$, Fig. \ref{fig:Rg2}(b), at {\it equal} shear rate $\dot{\gamma}$, the speed of knot tightening increases because the total drag force acting on the chain segments is proportionally larger. If, however, shear is expressed in Weissenberg numbers (which also account for larger relaxation times), this effect is reversed.}
%
%
\textcolor{black}{Averaging over several runs, one may determine the mean rate of tightening, $\dot{K}_n(t) = \frac{dK_n(t)}{dt}$, from the slope. In Fig. \ref{fig:Rg2}(c) this is shown for the case of a chain with $N=512$ by exploring the onset of non-zero tightening rate, $\dot{K}_n$. Notwithstanding the considerable fluctuations, one can observe a steady increase in $\dot{K}_n > 0$ beyond $\mathrm{Wi}^\mathrm{cr} \approx 7.97$, as indicated in Fig. \ref{fig:Rg2}(c).}

The transformation in time of the sheared coil conformation is indicated by the growth of the largest eigenvalue $\lambda_1$ of the tensor $\langle R_g^2(t)\rangle$ along with a simultaneous decline of the other two eigenvalues, $\lambda_2$ and $\lambda_3$. The original coil configuration thus turns into a stretched string while the total time it takes for this transformation steadily declines with increasing shear rate, cf. Fig.~\ref{fig:Rg2}(d), where one should note the logarithmic scale of the $Y$-axis.  


\begin{figure}[h!]
\includegraphics[width=.95\linewidth]{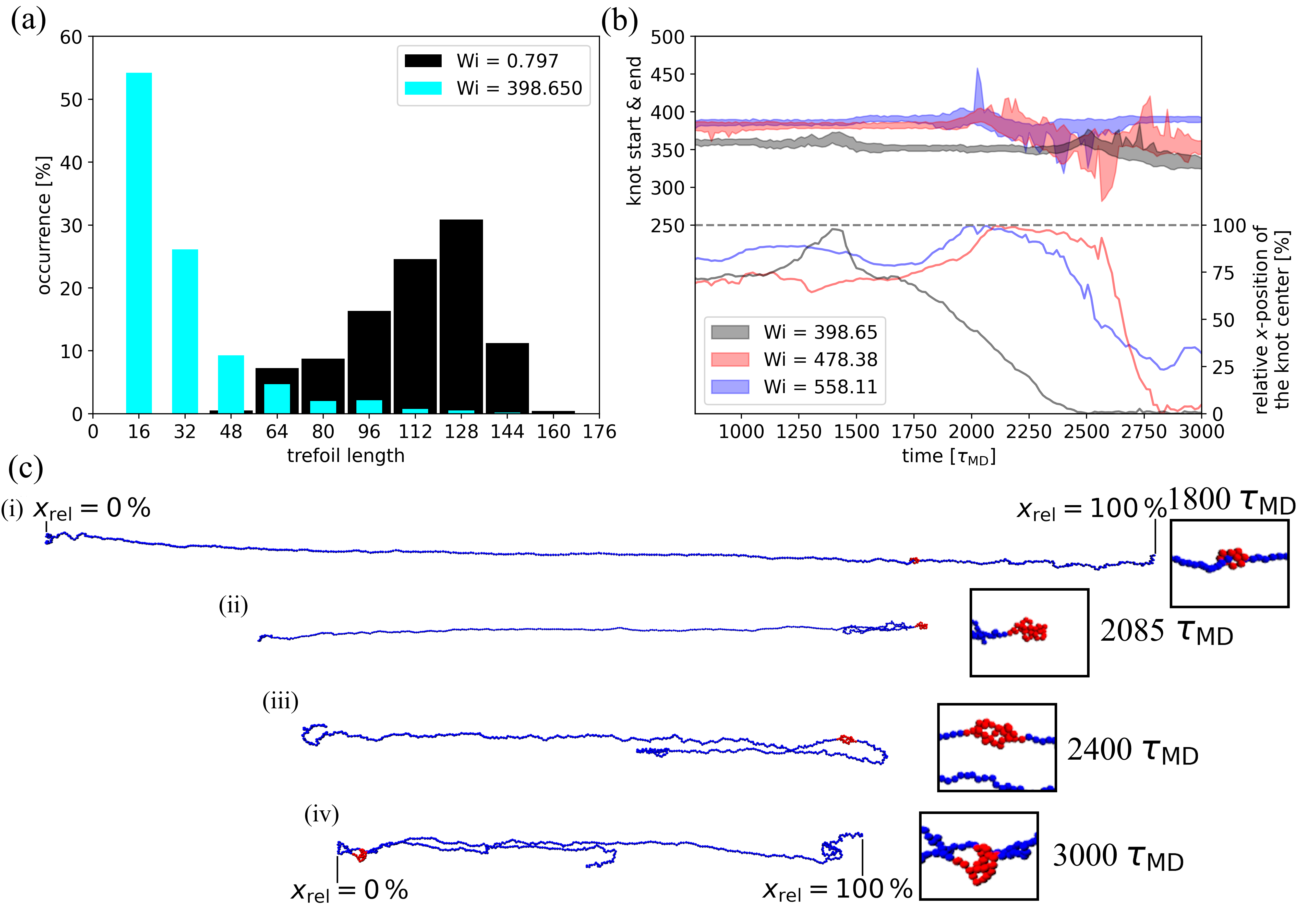}
\caption{\msrev{Tumbling effects for chains of size $N=512$. (a) Trefoil length distributions; only configurations after $t = 1000 \tau_\mathrm{MD}$ are analyzed to exclude the dynamic tightening phase. Tight knots are strongly favored at high Weissenberg numbers due to stretching Couette flow, but the distribution shows a long tail indicating that knots occasionally increase in size. (b) Kymographs of the knot start and end positions compared to the relative $x$-position of the knot center $x_\mathrm{rel}$ along the chain. Three individual simulation runs where tumbling occurred are shown: Knot sizes fluctuate and grow  whenever the knot center reaches the most extreme $x$-positions ($x_\mathrm{rel} \approx 100 \%$ or $x_\mathrm{rel} \approx 0 \%$). (c) Snapshots of the simulation run shown in (b) for Wi = 478.38, with 3x zooms onto the knotted regions. (i) The chain is completely stretched and the knot is in a non-diffusive tightened state. (ii,iii) The chain end with larger $x$-coordinate folds as tumbling begins, leaving the knot at $x_\mathrm{rel} \approx 100 \%$. At this point knot sizes start to fluctuate. (iv) The tumble is almost complete as the knot reaches $x_\mathrm{rel} \approx 0 \%$. }}
 \label{fig:tumbling_knotsizedistribution}
 \end{figure}

\textcolor{black}{In Figure \ref{fig:tumbling_knotsizedistribution} we investigate the complex and rich dynamics of knots in polymers of size $N=512$ after they reach a steady state. As observed in comparable simulations for (closed) knotted rings \cite{Liebetreu2018}, we obtain a broad distribution of knot sizes at low Weissenberg numbers indicating a weakly localized knot while knots are mostly small and tight for large shear rates. Intriguingly, just like in \cite{Liebetreu2018}, the latter distribution has a long tail, i.e., relatively large knots appear occasionally. The upper part of Fig.~\ref{fig:tumbling_knotsizedistribution}(b) displays a kymograph for three selected runs at large Weissenberg numbers which mark the start and end of the trefoil knot as the trajectory evolves. While the position of the knot along the chain is mostly fixed in the tight state, from time to time knots increase in size and become more mobile. These fluctuations are correlated with the relative position of the knot in shear direction. This position is 0 if the center bead of the knot is the monomer with the smallest x-coordinate of the chain and $100 \%$ if the center bead of the knot has the largest x-value. Close to these extreme values, fluctuations in knots sizes are most pronounced (lower part of Fig.~\ref{fig:tumbling_knotsizedistribution}(c)). As shown in the snapshots (Fig.~\ref{fig:tumbling_knotsizedistribution}(c)) for Wi=478.31, the chain undergoes a tumbling-like motion and folds back onto itself. Once the knot reaches a turning point, it loosens up and fluctuations in size persist for some time.} \msrev{A video of an individual simulation run exhibiting tumbling can be found in the supplementary media.}
\\
\textcolor{black}{Note that while this behavior is similar to the one observed in rings \cite{Liebetreu2018}, conformations in which the knot occupies half of the chain cannot be observed in our linear systems. These conformations arise in rings when all essential crossings of the knot are located in the center and the ring is effectively subdivided into two arcs of equal size.}


\subsection{Globule in Couette flow}
\begin{figure}[ht!]

\includegraphics[width=.95\linewidth]{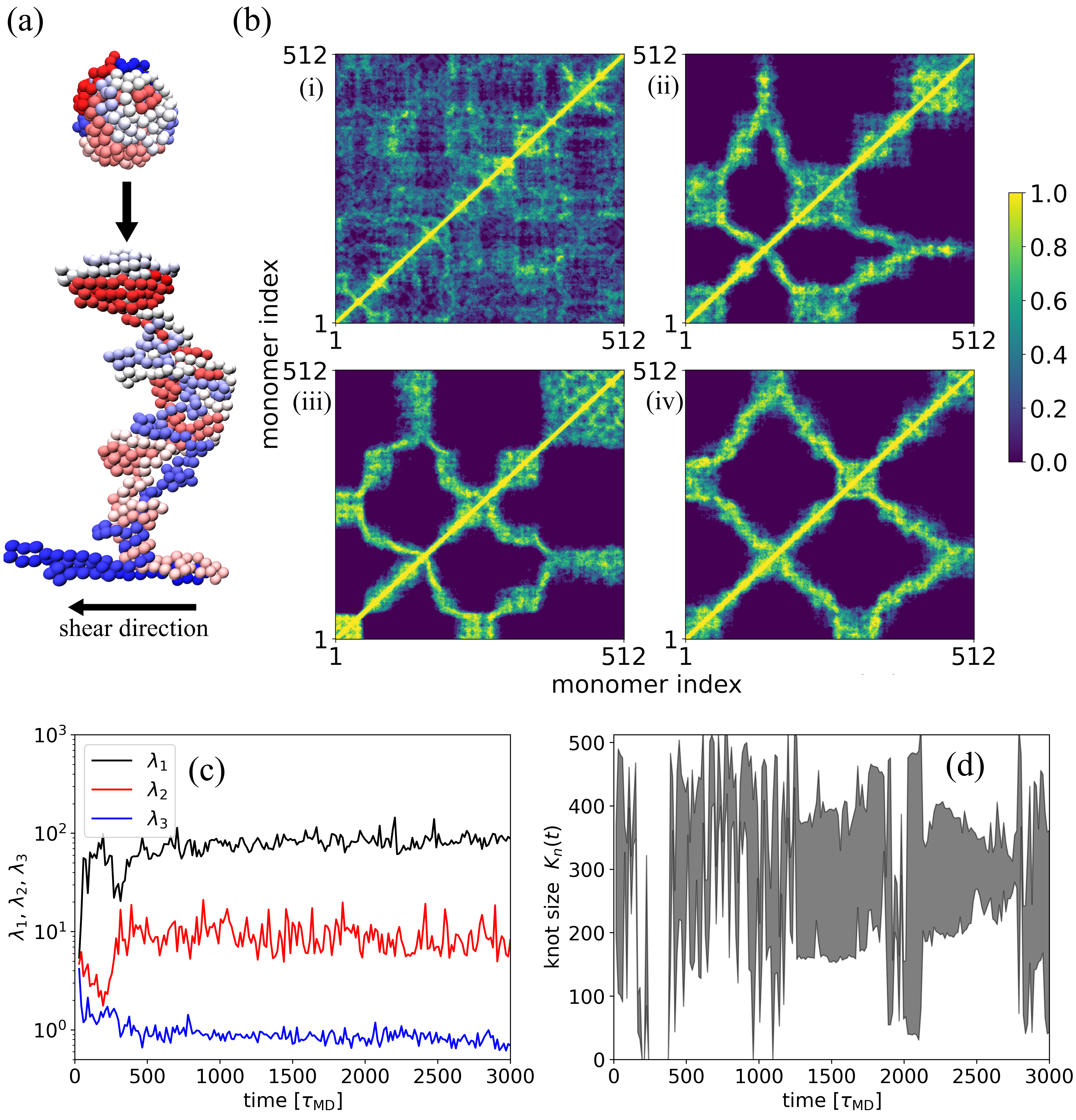}

\caption{\msrev{Structural and topological analysis of a representative simulation of a polymer globule with $N=512$ beads in Couette flow at $\dot\gamma = 0.75$. 
(a) Conformation of an initially globular polymer and a shear-induced pearl necklace configurations oriented along the $Y$-axis perpendicular to flow (corresponding to the contact map (b)iv). The continuous coloring scheme from red to white to blue indicates the monomer index. (b) Contact maps displaying monomers that come within $5 \sigma$ of other monomers, averaged over $5 \%$ of the simulation run starting at (i) $t = 0$, (ii) $t = 900 \tau_\mathrm{MD}$, (iii) $t = 1800 \tau_\mathrm{MD}$, and (iv) $t = 2850 \tau_\mathrm{MD}$. (c) Variation of the eigenvalues $\lambda_1, \lambda_2, \lambda_3$ of the tensor $R_g^2(t)$ with time indicating ongoing unfolding and refolding of the globule along the $Y$- (black), and $X$-axis (red). (d) Kymograph indicating the start and end of the knotted section in the course of the simulation run.}}

 \label{fig:KnV0.1glob}
 \end{figure}
 
In contrast to knots in coils, knots residing in globules are found to behave rather differently when subjected to shear. Earlier\cite{Kantor}, we demonstrated that knots in a globular configuration are much more loose and spread out as compared to those in coils. One should also bear in mind that about two orders of magnitude stronger shear rates, e.g.,  
$\dot{\gamma} = 0.75$, 
as those applied to coils are needed to unfold a globule, even partially, as reported earlier\cite{Netz} \msrev{($\dot{\gamma}=0.75$ would translate to $\mathrm{Wi}_{N=512}^\mathrm{coil}\approx 23919$ in the coil simulations, but we cannot measure the Weissenberg numbers for our globular simulations due to their very large relaxation times)}. Furthermore, at and beyond $\dot{\gamma} \approx 1.5$ \msrev{($\mathrm{Wi}_{N=512}^\mathrm{coil}\approx$ 47838)} our computer experiments indicate that a globule with $N=512$ beads eventually stretches into a linear string of beads {\it along} the flow (not shown here) albeit at such rates the bead-spring chain frequently breaks apart. 
 


At $\dot \gamma = 0.75$, the globule decays as a rule into several sub-globules which repeatedly merge and separate - see a typical snapshot \msrev{in Fig. \ref{fig:KnV0.1glob}(a)}. Note that a similar behavior characterizes the collapse of a polymer chain in a poor solvent whereby the intermediate stages of the folding mechanism include formation and coarsening of sub-globules as predicted by P.-G. de Gennes\cite{deGennes} and later confirmed in computer simulations\cite{Yeomans_2002,Ravi_2008,Janke_2017}. We find that the shear-induced unfolding of the globule \textcolor{black}{is somewhat reminiscent} of the {\it reverse} pathway of the collapse transition whereby the drag force of the shear is balanced by the attractive interaction of the chain segments. 
\textcolor{black}{However, there are also significant differences. Unlike the latter, remnants of the self-entangled globular starting conformation still persist in the pearl-necklace structure. The color scheme in Fig.~\ref{fig:KnV0.1glob}(a) (starting with red for the first bead, via white towards dark blue for the last) as well as the contact map for the late stages of the trajectory in Fig.~\ref{fig:KnV0.1glob}(b) (iv) clearly display that the polymer folds back onto itself several times and sub-globules are typically made up of non-consecutive sections. 
Contact maps for the late stages of eight more runs are shown in SI and differ significantly. Amongst those only the one in the center of Fig.~S2 can be identified as a "classical" pearl-necklace structure, i.e., a string of sub-globules which does not fold back onto itself.}


Interestingly, the globule stretches along the $Y$-axis, {\it perpendicular} to the direction of motion \textcolor{black}{(as indicated by the eigenvalue $\lambda_1$ of $\langle R_g^2(t)\rangle$ in Fig.~\ref{fig:KnV0.1glob}(c))}, $X$, while it remains almost flat in $Z$-direction \textcolor{black}{(as indicated by $\lambda_3$)}.
The resilience of globules against shear is also seen in Fig.~\ref{fig:KnV0.1glob}(c), where, in contrast to the case of sheared coils, one observes no rapid stretching as the eigenvalues of the $\langle R_g^2(t)\rangle$ tensor fluctuate strongly (note the logarithmic scale) around some constant \msrev{saturation} values during \msrev{most of the} 
time interval of the simulation. 

\textcolor{black}{The knotting behavior of this representative trajectory is analyzed qualitatively in Fig.~\ref{fig:KnV0.1glob}d. Here, a kymograph displays the start and the end of the knotted section over the course of the simulation run. Notably, knot sizes are often large, indicating that knots span multiple sub-globules, rather than being contained within a single one. Furthermore, the knot type also changes over the course of the representative simulation run. The initial trefoil knot first unties after about $10 \%$ of the total simulation time via one of the ends, reties to a trefoil knot after about $15 \%$, and then becomes rather complex after about $40 \%$ of the total simulation time (in places beyond the capabilities of the algorithm to assign a knot type).} \msrev{The supplementary media includes a video showing a globular simulation run.}

\section{Conclusions}

In this study we have demonstrated that shear flow leads to tightening of knots in linear polymer chains in good solvent conditions. The rate of knot tightening in polymer coils increases with growing shear rate\msrev{.} 
\textcolor{black}{Once the chain is tightened under strong shear, the polymer undergoes a tumbling-like motion which occasionally loosens up the knot.}
In bad solvent conditions, knotted polymer globules unroll partially under shear and attain a pearl-necklace conformation comprised of sub-globules somewhat reminiscent of intermediate states observed in the folding of globules. \textcolor{black}{Unlike the latter, the entangled polymer tends to fold back onto itself leading to a convoluted internal structure.}
These conformations stretch and orient themselves perpendicular to the flow direction and that of the velocity gradient $\dot{v}_z$. From a topological point of view, a complex knotting behavior can be observed in which knots are common and change rapidly.

Given that knots play an important yet not fully understood role in a wide range of macromolecular systems, ranging from DNA and proteins to synthetic polymers, we \textcolor{black}{hope that our findings on the fascinating behavior of knots under shear will become the foundation for further research efforts by computer simulations and single-molecule experiments.}



\section{Supplementary Material}
\textcolor{black}{See supplementary material for simulations and analyses of coils in Poiseuille flow and contact maps corresponding to the late stages of all performed globule simulations. Supplementary videos show the tumbling of a coil in Couette flow and a globule's transformation into a pearl-necklace configuration. }

\section{Acknowledgements}

M.P.S. and P.V. are grateful to the Deutsche Forschungsgemeinschaft (DFG) for funding (SFB TRR 146, project number \#233630050). A.M. and P.V. would also like to acknowledge partial support by the Deutsche Forschungsgemeinschaft (DFG) grant number VI $237/6-2$. The authors gratefully acknowledge the computing time granted on the supercomputer MOGON II and III at Johannes Gutenberg University Mainz as part of NHR South-West. 

\section{Author Declarations}
\subsection{Conflict of Interest}
The authors have no conflicts of interest to disclose.

\section{Data Availability Statement}
The data that support the findings of this study are available from the corresponding author
upon reasonable request.

\newpage
\bibliography{shear_knot.bib}


\end{document}